\documentclass[twocolumn,prl,superscriptaddress,showpacs]{revtex4}
\usepackage{amsmath}
\usepackage{amssymb}
\usepackage{bm}
\usepackage{epsfig}
\usepackage{graphicx}

\newcount\timehh  \newcount\timemm
\timehh=\time \divide\timehh by 60
\timemm=\time
\begin{document}

\title{Spin noise of exciton-polaritons in microcavities}

\author{M.M. Glazov}
\affiliation{Ioffe Physical-Technical Institute of the RAS, 194021 St.-Petersburg, Russia}
\affiliation{Spin and Optics Laboratory, St.-Petersburg State University, 198504
St.-Petersburg, Russia}
\author{M.A. Semina}
\affiliation{Ioffe Physical-Technical Institute of the RAS, 194021 St.-Petersburg, Russia}
\author{E.Ya. Sherman}
\affiliation{Department of Physical Chemistry, The University of the Basque Country,
48080 Bilbao, Spain}
\affiliation{IKERBASQUE Basque Foundation for Science, Bilbao, 48011 Bizkaia, Spain}
\author{A.V. Kavokin}
\affiliation{Spin and Optics Laboratory, St.-Petersburg State University, 198504
St.-Petersburg, Russia}
\affiliation{School of Physics and Astronomy, University of Southampton, Highfield,
Southampton SO17 1BJ, United Kingdom}

\begin{abstract}
We develop a theory of spin fluctuations of exciton-polaritons in a
semiconductor microcavity under the non-resonant unpolarized pumping. It is
shown that the corresponding spin noise is sensitive to the scattering rates in the system, occupation of the ground state, statistics of polaritons, and interactions. The spin noise spectrum drastically narrows in the polariton lasing regime due to formation of a polariton condensate, while its shape can become non-Lorentzian owing to interaction-induced spin decoherence.
\end{abstract}

\pacs{72.25.Rb, 72.70.+m, 71.36.+c, 03.75.Kk}
\maketitle



\emph{Introduction.} Quantum microcavity is a system where a semiconductor
quantum well is placed between the Bragg mirrors making it possible to
achieve the strong coupling between the light and matter. In this system the
energy is coherently transferred back and forth between the photon trapped
in the microcavity and the exciton, the elementary excitation of the
semiconductor. First observed by Weisbuch \textit{et al.}~\cite%
{PhysRevLett.69.3314}, the strong coupling results in the formation of mixed
light-matter quasiparticles, exciton-polaritons, extensively studied since
then~\cite{microcavities,sanvitto_timofeev}. The exciton-polaritons combine
an extremely small effective mass, inherited from the photon, and strong
interactions between themselves and with the environment due to the
excitonic fraction in this quasiparticle. These quasiparticles
are at the origin of several fascinating phenomena, which
primarily include polariton lasing~\cite{Christopoulos}: macroscopic accumulation of polaritons in a single quantum state (non-equilibrium
condensate) accompanied by spontaneous emission of coherent light by this
state.

In semiconductor microcavities exciton-polaritons are
characterized by the spin projections $s_{z}=\pm 1$ onto the structure
growth axis $z$ corresponding to the right or left circular polarization of
the photon and to the same spin component of the exciton. Superpositions of $s_{z}=\pm 1$ states give rise to the linear or elliptical
polarization of exciton-polaritons. Polariton lasers represent a model
bosonic system where the polarization of light emitted by a microcavity
directly corresponds to the spin state quasiparticles. This allows
studying the polariton spin dynamics by optical methods~\cite{PhysRevLett.89.077402}. A great amount of prominent spin-related effects
has been realized in microcavities, including self-induced
Larmor precession~\cite{PhysRevB.73.073303,solnyshkov07}, linear
polarization inversion~\cite{kavokin_inv}, optical spin Hall effect~\cite%
{kavokin05prl,Leyder:2007ve,Kammann:2012uq}, spin Meissner effect~\cite%
{Rubo2006227,PhysRevLett.105.256401,PhysRevLett.106.257401}, spin
multistability~\cite{PhysRevLett.98.236401,Paraiso:2010fk}, etc., see Refs.~
\cite{microcavities,0268-1242-25-1-013001} for reviews. Hence, quantum microcavities became a solid state playground to study interacting and, generally, non-equilibrium Bose systems.

In polariton lasers, the spontaneous symmetry breaking results in the
appearance of a stochastic vector polarization including circular
polarization related with the spin of a polariton condensate~
\cite{PhysRevLett.101.136409,PhysRevB.80.195309}. The Stokes vector
(pseudospin) of a condensate may be pinned to one of the crystal axes \cite
{Levrat}, due to the structure anisotropy. In this case the time-averaged
polarization of emission of polariton lasers is defined by pinning, while it momentary value fluctuates. The spin fluctuations give rise to the
``spin noise" which may be studied by optical measurements.
The spin fluctuations, being inherent to any system at equilibrium or
not, were first observed in 1981 in Na vapor~\cite{aleksandrov81}. 
The studies of spin fluctuations has become an important part of spintronics. 
The measurements of spin noise provide a crucial information on spin dynamics of carriers, excitons, and nuclei, 
including spin precession and relaxation rates and statistics of spin fluctuations. These characteristics are hardly accessible otherwise~\cite{Crooker_Noise,Mueller2010,crooker2010,1742-6596-324-1-012002,dahbashi:031906,crooker2012}.

Here we study theoretically the spin noise spectra of polariton lasers both
below and above the laser threshold and demonstrate that it is extremely
sensitive to the mean occupation number of the condensate and to the
statistics of polaritons.  Similar approach can be used for a variety of systems where condensates of spin polarized bosons interact with an incoherent reservoir. An appropriate example is given by magnon condensates in crystals, either strongly driven by periodic fields~\cite{demokritov} or imposed to sufficiently strong static magnetic field at a low temperature~\cite{giamarchi}. Here the noise
is seen as fluctuations in the sample magnetization. The finite-temperature Bose-Einstein condensates of atoms in optical traps with the synthetic spin-orbit coupling represent another important example of direct applicability of our approach~\cite{Lin}.


\emph{Model.} The dynamics of the exciton-polariton spin doublet can be 
described with the pseudospin approach where the density matrix of polariton
state with the wavevector $\bm k$ can be written as $\hat{\rho}_{\bm k}=N_{%
\bm k}\hat{I}+\bm S_{\bm k}\cdot \hat{\bm\sigma }$. Here $\hat{I}$ is the
unit $2\times 2$ matrix, $\hat{\bm\sigma }$ is a pseudovector composed of
Pauli matrices, $N_{\bm k}$ is the spin-average occupancy of the state $\bm k
$, and $\bm S_{\bm k}$ is the pseudospin of polaritons in this state. In what
follows we focus on the dynamics of the ground state, corresponding to $\bm %
k=0$, treat all other states as a reservoir, and omit $\bm k$ subscript in the notations. The pseudospin components $S_{\alpha }$, where $\alpha =\{x,y,z\}$ is the
Cartesian index corresponding to the polarization of 
emission: $S_{z}/N$ gives its circular polarization degree, while $S_{x}/N$
and $S_{y}/N$ give the linear polarization degree in the axes frames $(xy)$
and $(x^{\prime }y^{\prime })$ rotated by $45^{\circ }$ with respect to each
other.~\cite{microcavities} The dynamics of the occupation number and the
polariton pseudospin is governed by the set of kinetic equations~\cite{0268-1242-25-1-013001} 
\begin{eqnarray}
&&\frac{dN}{dt}+\frac{N}{\tau _{0}}+Q_{n}\{N,\bm S\}=0,  \label{eqN} \\
&&\frac{d\bm S}{dt}+\bm S\times \bm\Omega +\frac{\bm S}{\tau _{0}}+{\bm Q}%
_{s}\{\bm S,N\}=0,  \label{eqS}
\end{eqnarray}%
where $\tau _{0}$ is the lifetime of polaritons in the ground state, $\bm%
\Omega $ is the effective magnetic field arising from anisotropy of the
system and from polariton-polariton interactions, its explicit form
will be given below. The scalar, $Q_{n}\{N,\bm S\}$, and pseudovector, ${\bm Q}_{s}\{\bm S,N\}$, collision integrals describe arrival and departure of
the particles into the ground state, see Refs.~\cite{0268-1242-25-1-013001,PhysRevB.73.073303} for general
expressions. Here we adopt their simplest form 
\begin{eqnarray}
&&Q_{n}\{N,\bm S\}=W^{out}N-W^{in}(1+N),  \label{Qn} \\
&&{\bm Q}_{s}\{\bm S,N\}=(W^{out}-W^{in}+\tau _{s}^{-1})[\bm S-\bm S_{0}(\bm%
\Omega )].  \label{Qs}
\end{eqnarray}%
Here $W^{out}$ and $W^{in}$ are out- and in-scattering rates, related with
the presence of the reservoir, as schematically illustrated in Fig.~\ref%
{fig:pumping}(a). In particular, $W^{in}$ is proportional to the occupation of
the reservoir and it is determined by the pumping rate. The term
proportional to $1+N$  describes the stimulated transitions due to the bosonic statistics 
of quasiparticles. In the collision integral
for the polariton spin, $\bm Q_{s}\{\bm S,N\}$, $\tau _{s}$ is the spin
relaxation time and $\bm S_{0}(\bm\Omega )$ is the steady-state spin induced
by the effective field $\bm\Omega $. If polariton exchange with reservoir is
efficient, one can introduce the effective temperature $T$ of the polariton
system. The steady-state spin is $-\langle N\rangle {\bm\Omega }/|\bm\Omega |
$ for $T\ll \Omega $ and $0$ for $T\gg \Omega $. In what follows, unless
otherwise specified, we consider the case of high temperatures, where $%
\langle \bm S\rangle \equiv 0$, and the occupancy of the polariton ground
state given by the balance of in- and out- scattering processes is: 
\begin{equation}
\langle N\rangle =\frac{W^{in}}{\tau _{0}^{-1}+W^{out}-W^{in}},
\label{steady}
\end{equation}%
where the condition $W^{in}<\tau _{0}^{-1}+W^{out}$ should hold~\cite{microexpr}. 
The straightforward generalization of our approach to
account for the entire ensemble of polaritons following
Refs.~\cite{0268-1242-25-1-013001,PhysRevB.73.073303} using the full
density matrix is not expected to yield qualitatively different results.

\begin{figure}[t]
\includegraphics[width=\linewidth]{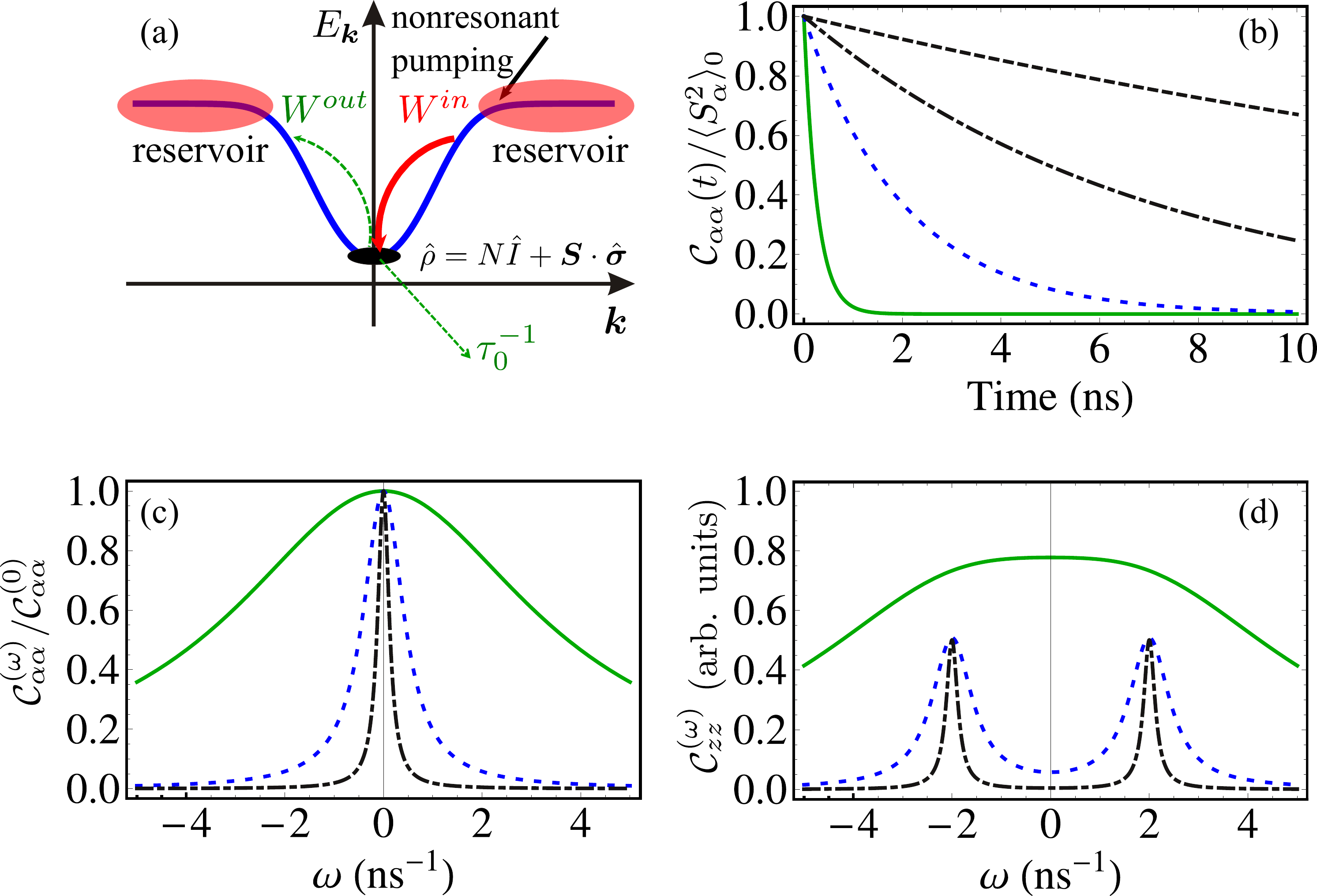} 
\caption{(a) Illustrative scheme of the pumping. The reservoir and the
ground state are shown as well as in- and out- scattering processes. (b)
Temporal dependence of spin fluctuations calculated for the $\langle
N\rangle = 10$ (green/solid), $100$ (blue/dotted) $1000$
(black/dash-dotted). Other parameters are: $\protect\tau_0 = 25$~ps, $%
W^{out}=0$, $\protect\tau_s = 10$~ns. Dashed curve is calculated for $%
\protect\tau_s\to \infty$ and $\langle N\rangle = 1000$. (c) Spin noise
power spectra calculated for the same parameters. (d) Spin noise power
spectra calculated with allowance for the anisotropic splitting $\Omega_a =2$%
~ns$^{-1}$. Other parameters are the same as in panels (b), (c).}
\label{fig:pumping}
\end{figure}

The fluctuations of the condensate occupation number $\delta N(t)\equiv
N(t)-\langle N\rangle $, and pseudospin $\delta \bm S(t)\equiv \bm %
S(t)-\langle \bm S\rangle =\bm S(t)$ are described by the correlation
functions, namely: $\mathcal{K}(t)\equiv \langle \delta N(t^{\prime })\delta
N(t^{\prime }+t)\rangle $ and $\mathcal{C}_{\alpha \beta }(t)\equiv \langle
S_{\alpha }(t^{\prime })S_{\beta }(t^{\prime }+t)\rangle $, where the
angular brackets denote the averaging over the time $t^{\prime }$ for a
given shift $t$. According to the general theory of fluctuations~\cite%
{springerlink:10.1007/BF02724353,ll10_eng,ivchenko73fluct_eng} their
correlation functions obey the same set of kinetic equations for $t$ or $t^{\prime}-$dependence as fluctuating quantities~\cite%
{noiselinear}. The solution of Eqs.~\eqref{eqN} and \eqref{eqS} in the
absence of effective magnetic fields and interactions ($\bm\Omega \equiv 0$%
), results in the exponential time-decay of correlations and isotropic spin
fluctuations: 
\begin{equation}
\mathcal{K}(t)=\mathcal{K}(0)e^{-{|t|}/{\tau _{\mathrm{c}}}},\quad \mathcal{C%
}_{\alpha \beta }(t)=\delta _{\alpha \beta }\mathcal{C}_{\alpha \alpha
}(0)e^{-{|t|}/{\tau _{c,s}}},  \label{noise}
\end{equation}%
where $\delta _{\alpha \beta }$ is the Kronecker $\delta $-symbol, single
time correlators (mean square fluctuations) $\mathcal{K}(0)=\langle (\delta
N)^{2}\rangle $, $\mathcal{C}_{\alpha \alpha }(0)=\langle S_{\alpha
}^{2}\rangle _{0}$ will be found below, while the particle-spin correlations 
$\langle \delta N(t^{\prime })S_{\alpha }(t^{\prime }+t)\rangle $ vanish and
will be disregarded. We introduced the correlation times $\tau _{\mathrm{c}}$%
, $\tau _{c,s}$ according to 
\begin{equation}
\frac{1}{\tau _{\mathrm{c}}}=\frac{1}{\tau _{0}}+W^{out}-W^{in}=\frac{\tau
_{0}^{-1}+W^{out}}{1+\langle N\rangle },\quad \frac{1}{\tau _{c,s}}=\frac{1}{%
\tau _{\mathrm{c}}}+\frac{1}{\tau _{s}}.  \label{tauc}
\end{equation}%
Equations~\eqref{noise} and \eqref{tauc} clearly show that the particle
number and spin fluctuations of exciton-polaritons decay exponentially with 
time and the correlation time of the particle
number fluctuations is $\tau _{\mathrm{c}}$, while the spin fluctuations vanish faster, at
$\tau_{c,s}<\tau_{\mathrm{c}}$. The spin fluctuation
spectra defined as $\mathcal{C}_{\alpha \beta }^{(\omega )}\equiv
\int_{-\infty }^{\infty }\mathcal{C}_{\alpha \beta }(t)e^{i\omega t}dt$  are Lorentzian: 
\begin{equation}
\mathcal{C}_{\alpha \alpha }^{(\omega )}=\mathcal{C}_{\alpha \alpha }(0)%
\frac{2\tau _{c,s}}{1+\omega ^{2}\tau _{c,s}^{2}},  \label{sns}
\end{equation}%
with the half-width half maximum $\tau _{c,s}^{-1}$ determined by the
inverse spin correlation time. It follows from Eq.~\eqref{tauc} that the
correlation time of fluctuations is strongly enhanced and the spin noise
spectrum is strongly narrowed, if $\langle N\rangle \gg 1$, i.e. where the
ground state is macroscopically occupied. In particular, if $\tau_{s}\rightarrow \infty $ and $W^{out}\rightarrow 0$ which corresponds to the
negligible spin-flip in the ground state and negligible depletion of the
condensate due to the scattering of polaritons back to the reservoir, we
have for the correlation time of particle and spin fluctuations $\tau
_{c}=\tau _{0}(1+\langle N\rangle )$. The life-time of exciton-polaritons in
the state-of-the-art structures varies from $\sim 1$~ps to $\sim 100$~ps
with the corresponding maximum $\langle N\rangle $ being between $10^{3}$
and $10^{5}$, yielding $\tau _{c}$ in the range of $1$~ns \ldots $10$~$\mu $s. Hence, the spin noise frequencies range from MHz to GHz for the
macroscopically occupied ground state. The typical spin noise spectra and
temporal dependence of the correlators in Fig.~\ref%
{fig:pumping}(b),(c) make the drastic effect of the ground state occupation
clear. The several orders of magnitude enhancement of the spin
correlation time and narrowing of the noise spectrum is a general bosonic
effect. Indeed, any fluctuation is amplified by the bosonic stimulation, factor $1+N$ in Eq.~\eqref{Qn}.

The mean square of the particle and pseudospin fluctuations can be found
using the master equation approach~
\cite{sanvitto_timofeev}. In the absence of interactions and effective magnetic
fields, the system is spin-isotropic and can be described by
the independent occupations of two orthogonal spin states, $N_{\uparrow }$
and $N_{\downarrow }$, described by the same distribution functions $%
P(N_{\uparrow ,\downarrow })$. Using these distribution functions one can
express the mean square fluctuations, that is $t=0$ correlators $\langle
(\delta N)^{2}\rangle =\sum_{N_{\uparrow },N_{\downarrow }}P(N_{\uparrow
})P(N_{\downarrow })[(N_{\uparrow }+N_{\downarrow })/2-\langle N\rangle ]^{2}
$ and $\langle S_{\alpha }^{2}\rangle _{0}=\sum_{N_{\uparrow },N_{\downarrow
}}P(N_{\uparrow })P(N_{\downarrow })[(N_{\uparrow }-N_{\downarrow })/2]^{2}$%
, as 
\begin{equation}
\langle S_{\alpha }^{2}\rangle _{0}=\langle (\delta N)^{2}\rangle =\frac{1}{2%
}\langle N\rangle \lbrack 1+(g^{(2)}-1)\langle N\rangle ],  \label{g2}
\end{equation}%
where $\langle S_{\alpha }^{2}\rangle _{0}$ corresponds to isotropic
fluctuations, and $g^{(2)}$ is the second order coherence of a single state.
In particular, $g^{(2)}=2$ corresponds to the thermal statistics~\cite%
{Glauber}, where particle and spin square fluctuations are $\propto \langle
N\rangle (1+\langle N\rangle )$ and grow quadratically with the ground state
occupation. This situation is realized for equilibrium Bose gas~\cite{ll5_eng} 
or if the ground state feedback on reservoir is negligible. By
contrast, $g^{(2)}=1$ corresponds to the coherent statistics, where the mean
square fluctuations are suppressed, being $\propto \langle N\rangle $. In
the limit of  low occupancy, $\langle (\delta N)^{2}\rangle=\langle N\rangle$ 
in agreement with the theory of classical gas. Moreover, the statistics of spin fluctuations can be
determined, as a convolution of distribution functions $P(N-S_{\alpha })$
and $P(N+S_{\alpha })$. In the limit $\langle N\rangle \gg 1$ we obtain: 
\begin{eqnarray}
&&\hspace{-0.5cm}p_{\mathrm{coh}}(S_{\alpha })=(\pi \langle N\rangle
)^{-1/2}\exp {(-S_{\alpha }^{2}/\langle N\rangle )},~~g^{(2)}=1,
\label{distribs_coh} \\
&&\hspace{-0.5cm}p_{\mathrm{th}}(S_{\alpha })=\langle N\rangle ^{-1}\exp {%
(-2|S_{\alpha }|/\langle N\rangle )},~~g^{(2)}=2.  \label{distribs_th}
\end{eqnarray}%
%
The full statistics of polariton condensates can be determined by numerical integration of
Langevin equations for the condensate wavefunctions, as it was done, e.g. in Ref.~\cite{PhysRevB.80.195309}
for stochastic polarization under pulsed excitation. Here we resort to an analytical approach based on kinetic equations
treating $g^{(2)}$ phenomenologically.

\emph{Role of effective magnetic fields.} We begin the discussion of the
effective magnetic fields with the case of an anisotropic system where the
polariton doublet is split into the pair of states, linearly polarized along 
$x$ and $y$ axes. In such a case the vector $\bm\Omega =(\Omega _{a},0,0)$
determines the anisotropic splitting~\cite%
{Klopotowski2006511,PhysRevB.73.073303,amo09,PhysRevB.82.085315}, and we
assume that the effective temperature $T$ of the system exceeds the
anisotropic splitting to neglect the steady spin polarization. The $S_{z}$
and $S_{y}$ rotate, while the dynamics of $S_{x}$ remains purely
dissipational. As a result, the fluctuations become anisotropic with nonzero
spectrum components (c.f. Ref.~\cite{gi2012noise}) 
\begin{eqnarray}
&&\hspace{-0.7cm}\frac{\mathcal{C}_{xx}^{(\omega )}}{\langle S_{\alpha
}^{2}\rangle _{0}}=\frac{2\tau _{c,s}}{1+\omega ^{2}\tau _{c,s}^{2}},~\frac{%
\mathcal{C}_{yy}^{(\omega )}}{\langle S_{\alpha }^{2}\rangle _{0}}=\sum_{\pm}%
\frac{\tau _{c,s}}{1+(\omega \pm\Omega _{a})^{2}\tau _{c,s}^{2}},
\label{sns:a:diag} \\
&&\hspace{-0.7cm}\mathcal{C}_{yz}^{(\omega )}=\frac{2i\omega \Omega _{a}\tau
_{\mathrm{c},s}^{2}}{1+\tau _{c,s}^{2}(\omega ^{2}+\Omega _{a}^{2})}\mathcal{%
C}_{yy}^{(\omega )},  \label{sns:a:ndiag}
\end{eqnarray}%
and $\mathcal{C}_{zz}^{(\omega )}=\mathcal{C}%
_{yy}^{(\omega )}$, $\mathcal{C}_{yz}^{(\omega )}=[ \mathcal{C}%
_{zy}^{(\omega )}] ^{\ast }$, $\langle S_{\alpha }^{2}\rangle _{0}$ is given
by Eq.~\eqref{g2}, and $\tau _{c,s}$ by Eq.~\eqref{tauc}. Figure~\ref%
{fig:pumping}(d) shows the spin noise spectra calculated with the allowance
for the anisotropic splitting for different occupations of the ground state.
It is seen that the single peak is transformed into the two peak structure even for a very small value of $\hbar\Omega_a$ taken in
our calculation. This is because the spin correlation time $\tau _{c,s}$
depends strongly on the ground state occupation: for small pumping rates and
small $\langle N\rangle $ the product $\Omega _{a}\tau _{c,s}\ll 1$ and the
splitting is not visible, however, for larger pumping $\Omega _{a}\tau _{c,s}
$ becomes comparable or larger than 1, making the anisotropic splitting 
$\hbar \Omega _{a}$  resolvable, as demonstrated in Fig.~\ref%
{fig:pumping}(d). If the temperature is so low that $T\ll \hbar \Omega _{a}$%
, the polaritons are predominantly polarized along the effective field
direction $x$. In this case, $\langle S_{x}\rangle =-\langle N\rangle $, $%
\langle S_{y}\rangle =\langle S_{z}\rangle =0$, while the mean square
fluctuations take the form $\langle (S_{x}-\langle S_{x}\rangle )^{2}\rangle
=\langle (\delta N)^{2}\rangle $, $\langle S_{y}^{2}\rangle =\langle
S_{z}^{2}\rangle =\langle N\rangle /2$. Moreover, the non-trivial single
time correlation appears: $\langle S_{y}S_{z}\rangle =\langle
S_{z}S_{y}\rangle ^{\ast }=-i\langle N\rangle /2$. Spin noise spectrum in
this case can be obtained in a similar way. Equation \eqref{sns:a:diag} holds here, albeit with different mean square fluctuations 
found above, in the numerators.

Now let us consider the effect of polariton-polariton interactions on the
spin noise. As follows from multiple experimental and theoretical works~\cite%
{kavokin_inv,PhysRevB.80.155306,Rubo2006227,PhysRevLett.105.256401,PhysRevLett.106.257401,Paraiso:2010fk}
these interactions are strongly spin-anisotropic: the exciton-polaritons
with the same $z$ pseudospin components, i.e. with the same circular
polarizations, repel each other efficiently due to the exchange interaction of
electrons/holes with the same spin, while the polaritons with opposite
circular polarizations can weakly attract each other, the latter is neglected.  
Hence, the interactions create an effective fluctuating field $\bm\Omega
=(0,0,\Omega _{i})$, directed along the $z$ axis. Its magnitude $\hbar
\Omega _{i}\equiv \alpha _{1}S_{z}$ ($\alpha _{1}>0$ is
the interaction constant) is related to the
fluctuations of polaritons pseudospin $z$ component. This effective field
has two important effects on polariton spin dynamics and spin noise: (i) it
induces precession of the pseudospin around the $z$ axis, known as
self-induced Larmor precession~\cite{PhysRevB.73.073303,solnyshkov07}, and
(ii) it suppresses fluctuations of the $z$ pseudospin component, favoring
linear polarization of the macrooccupied state~\cite{Rubo2006227,laussy2006}.

It is instructive to start the analysis with the spin precession effect
assuming that the effective temperature is high enough, $T \gg \alpha_1 
\langle S_z^2\rangle_{0}^{1/2}$. In addition, we neglect below the
anisotropic splitting ($\Omega_a=0$). Clearly, the effective field $%
\Omega_i$ induces dephasing of the in-plane pseudospin components.
It follows from Eq.~\eqref{eqS} that %
\begin{equation}  \label{inplane}
\frac{ d \mathcal{C}_{\alpha\beta}(t)}{ dt} + \frac{\mathcal{C}%
_{\alpha\beta}(t)}{\tau_{ c,s}} {\pm} \frac{\alpha_{1}}{\hbar} \mathcal{C}%
_{\alpha\beta}(t) S_z(t) = 0.
\end{equation}
Here the upper (lower) sign corresponds to $\alpha=x$ $(y)$, correlations
between $z-$ and in-plane components are disregarded, since the field $\bm %
\Omega$ does not couple them, and $S_z$ can be considered as an independent
parameter whose fluctuations are given by Eqs.~\eqref{noise}, \eqref{sns}.
The set of linear Eqs~\eqref{inplane} can be readily solved as: 
\begin{equation}  \label{corr:int:gen}
\frac{\mathcal{C}_{\alpha\alpha}(t)}{\langle S_\alpha^2\rangle_{0}} = e^{-{%
|t|}/{\tau_{ c,s}}} \left\langle\exp{\left[ i \frac{\alpha_1}{\hbar}
\int_0^{|t|} S_z(t_1) dt_1 \right]}\right\rangle_z,
\end{equation}
with $\langle \ldots \rangle_z$ meaning the averaging over the fluctuations
of $S_z$. The treatment of the general case is beyond the scope of this
work, here we consider two limiting cases: the regimes of fast and slow
fluctuations, respectively.

If $\alpha _{1}\langle S_{z}^{2}\rangle _{0}^{1/2}\tau _{c,s}/\hbar \ll 1$,
the interaction induced effective field changes much faster than the psuedospin rotates. This case corresponds to the motional narrowing and 
\begin{equation}
\frac{\mathcal{C}_{\alpha \alpha }(t)}{\langle S_{\alpha }^{2}\rangle _{0}}%
=\exp {\left( -\frac{|t|}{\tau _{c,s}}-\frac{\alpha _{1}^{2}}{\hbar ^{2}}%
\langle S_{\alpha }^{2}\rangle _{0}\tau _{c,s}|t|\right) }.
\label{corr:fast}
\end{equation}%
In this limit the spin fluctuations decay exponentially, resulting in the
Lorentzian spectrum of spin noise. This spectrum is,
however, anisotropic: the width of $\mathcal{C}
_{xx}^{(\omega )}$ and $\mathcal{C}_{yy}^{(\omega )}$ is larger than that of 
$\mathcal{C}_{zz}^{(\omega )}$ because the interaction-induced field ${\bm \Omega}\parallel z$ does
not affect $S_{z}$.

\begin{figure}[t]
\includegraphics[width=\linewidth]{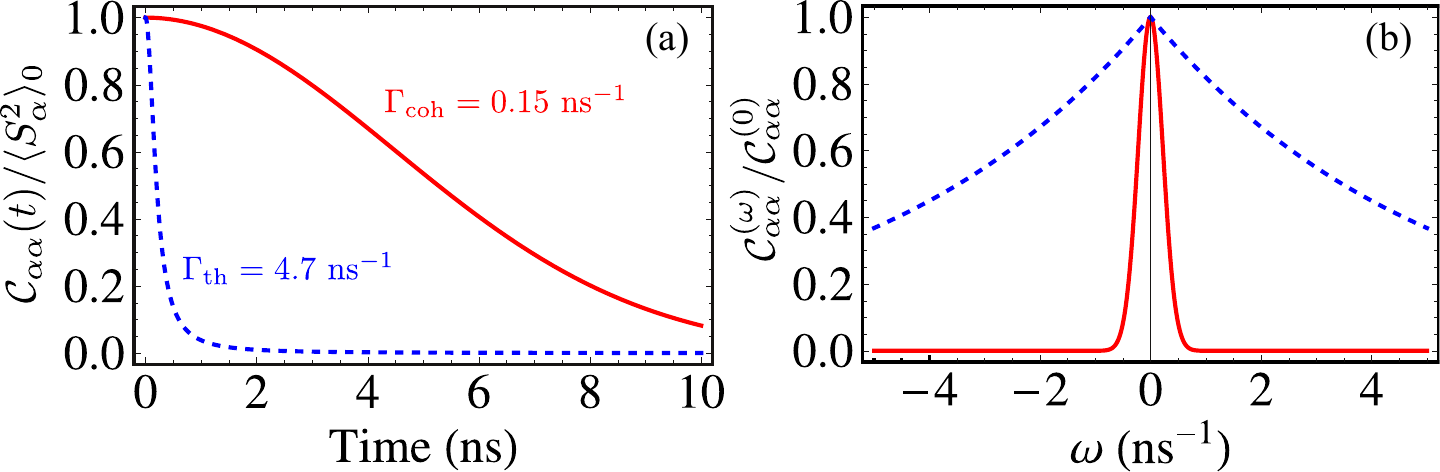}
\caption{(a) Temporal dependence of the in-plane pseudospin fluctuations
calculated with allowance for interactions according to Eqs.~ 
\eqref{strong:int1} and \eqref{strong:int2} for $\langle N \rangle = 1000$, $%
\protect\alpha_1 =10^{-2}$~ns$^{-1}$~\protect\cite{PhysRevB.80.155306} for
coherent statistics (red/solid) and thermal statistics (blue/dotted). (b)
Spin noise power spectra calculated for the same parameters as in (a).}
\label{fig:fluctTC}
\end{figure}

In the opposite limit, where the fluctuations of spin $z$ component are slow
enough and the in-plane spin components make several oscillations during the
correlation time of $S_{z}$, i.e. $\alpha _{1}\langle S_{z}^{2}\rangle
_{0}^{1/2}\tau _{c,s}/\hbar \gg 1$, the fluctuations of $S_{z}$ and of the
effective field $\Omega _{i}$ can be assumed frozen. In this case, the
spin dephasing takes place on the timescale of $\Omega _{i}^{-1}$. The
particular $t-$dependence of the correlator $\mathcal{C}_{\alpha \alpha }(t)$
is determined by statistics of the condensate~\cite{ivplaussy}. For $\langle
N\rangle \gg 1$ we obtain for two important limiting cases of coherent and
thermal statistics ($t\ll \tau _{\mathrm{c,s}}$): 
\begin{eqnarray}
&&\frac{\mathcal{C}_{\alpha \alpha }(t)}{\langle S_{\alpha }^{2}\rangle _{0}}%
=\exp {\left( -\Gamma _{\mathrm{coh}}^{2}t^{2}\right) },\quad g^{(2)}=1,
\label{strong:int1} \\
&&\frac{\mathcal{C}_{\alpha \alpha }(t)}{\langle S_{\alpha }^{2}\rangle _{0}}%
=\frac{1}{1+\Gamma _{\mathrm{th}}^{2}t^{2}},\quad g^{(2)}=2,
\label{strong:int2}
\end{eqnarray}%
where the dephasing rates are: $\Gamma _{\mathrm{coh}}^{2}=\alpha
_{1}^{2}\langle N\rangle /4\hbar ^{2}$ and $\Gamma _{\mathrm{th}}^{2}=\alpha
_{1}^{2}\langle N\rangle ^{2}/4\hbar ^{2}=\langle N\rangle \Gamma _{\mathrm{%
coh}}^{2}$. In this limit, the temporal dependence of the spin fluctuations
is directly related to the ground state statistics: Gaussian fluctuations of 
$S_{z}$ described by $p_{\mathrm{coh}}(S_{z})$ in Eq.~\eqref{distribs_coh}
result in the Gaussian decay of the in-plane spin components, while the
sharply-peaked $p_{\mathrm{th}}(S_{z})$ in Eq.~%
\eqref{distribs_th} for thermal statistics results in the slow power-law
decay of the fluctuations due to high probability of small $S_{z}$, 
corresponding to low precession rates. As a result, the noise spectrum of
the in-plane pseudospin components deviates strongly from Lorentzian. Doing Fourier transform of Eqs.~\eqref{strong:int1}, \eqref{strong:int2} we
obtain: 
\begin{eqnarray}
&&\frac{\mathcal{C}_{\alpha \alpha }^{(\omega )}}{\langle S_{\alpha
}^{2}\rangle _{0}}=\frac{\sqrt{\pi }}{\Gamma _{\mathrm{coh}}}\exp {\left(
-\omega ^{2}/4\Gamma _{\mathrm{coh}}^{2}\right) },\quad g^{(2)}=1,
\label{strong:int:sns1} \\
&&\frac{\mathcal{C}_{\alpha \alpha }^{(\omega )}}{\langle S_{\alpha
}^{2}\rangle _{0}}=\frac{\pi }{\Gamma _{\mathrm{th}}}\exp {\left( -|\omega
|/\Gamma _{\mathrm{th}}\right) },\quad g^{(2)}=2.  \label{strong:int:sns2}
\end{eqnarray}%
The calculated non-analytical dependence of the spin
fluctuations for the thermal statistics at $\omega \rightarrow 0$ is related
to the power-law temporal decay of the fluctuations. The typical temporal
dependence of the in-plane pseudospin correlators and noise power spectra
are plotted in Fig.~\ref{fig:fluctTC}(a) and (b), respectively. Figure
clearly shows different qualitative behavior of the noise of the in-plane
pseudospin components for different statistics of the polaritons. We stress
that the drastic difference of the decoherence times and noise spectral
widths for coherent and thermal statistics is related to different
dependences of $\langle S_{z}^{2}\rangle $ on the ground state occupancy, $%
\langle N\rangle $, see Eq.~\eqref{g2}.

Note, that if $\alpha _{1}\langle S_{z}^{2}\rangle $ is comparable with or
larger than the effective temperature $T$ of the system, the fluctuations of
the pseudospin $z$ component would be suppressed. This effect can be modeled
by multiplying the distribution function of $S_{z}$ in Eqs.~%
\eqref{distribs_coh} and \eqref{distribs_th} by the Boltzmann factor $\exp {%
(-\hbar \alpha _{1}S_{z}^{2}/T)}$ for the probability of thermal
fluctuations~\cite{Rubo2006227}. As a result, in the limit of $T\rightarrow
0 $ we obtain $\langle S_{z}^{2}\rangle =T/2\hbar \alpha _{1}$. In this
case, the dephasing rate, which determines the spin noise spectral width, can be estimated as $\sim \sqrt{\alpha _{1}T/\hbar }$.

\emph{Conclusions.} We have developed an analytical theory of spin fluctuations of polaritons 
in microcavities in the lasing regime and demonstrated that 
the spin noise spectra, being a fundamental property of 
any spin system,  qualitatively depend on the occupation numbers, statistics, and interactions
between the particles.  Various regimes of spin noise have been
identified.  Thus, spin noise spectroscopy allows one to study a large variety of 
spin-related properties of bosonic systems in a single experiment. Experimental verification of these predictions can be done by
Fourier spectroscopy of Kerr or Faraday rotation. Extension of this
model to other spin-polarized bosonic systems is straightforward. 

\emph{Acknowledgments.} The financial support from the Russian Ministry of
Education and Science (contract No. 11.G34.31.0067 with SPbSU and leading
scientist A. V. Kavokin), RFBR, RF President Grants NSh-5442.2012.2 and
2901.2012.2, EU project POLAPHEN is acknowledged. EYS was supported by the
University of Basque Country UPV/EHU program UFI 11/55, Spanish MEC
(FIS2012-36673-C03-01), and ``Grupos Consolidados UPV/EHU del Gobierno
Vasco'' (IT-472-10).


\begin{thebibliography}
\bibitem{microexpr} The analysis of the microscopic expressions for $W^{in}$
and $W^{out}$ shows that this condition is indeed fulfilled.
\bibitem{noiselinear} Generally, equations and collision integrals
should be linearized, but Eqs.~\eqref{Qn},\eqref{Qs} are already linear in $%
\bm S$ and $\delta N$.
\end{thebibliography}
\end{document}